\documentclass[twocolumn,showpacs,preprintnumbers,amsmath,amssymb,pre]{revtex4}
\usepackage{graphicx}
\usepackage{color}

\begin{document}
\preprint{}

\title{Foams in a rotating drum}

\author{A.~Bronfort}
\email{abronfort@ulg.ac.be}
\author{H.~Caps}

\affiliation{GRASP, Physics Department B5, University of Li\`ege, B-4000 Li\`ege, Belgium.}

\date{\today}

\begin{abstract}
A nearly two-dimensionnal foam is generated inside a Hele-Shaw cell half-filled with a surfactant solution. The cell is then placed vertically on a tumbler so that it rotates around its center with the cell angular velocity as the control parameter. During the foam rotation the liquid fraction increases and the foam/liquid interface deforms. The shear velocity profiles generated inside the foam are studied along two representative lines. The foam rheological properties are obtained using statistical tools and allow us  to determine the foam effective viscosity. A semiempirical model is proposed for these velocity profiles, emphasizing the significance of the viscous dissipations between the bubbles and with the cell walls.
\end{abstract}

\maketitle

\section{Introduction} 					

Liquid foams are complex fluids made of gas bubbles surrounded by a complex network of liquid channels. They are used in our everyday life as well as in many industrial applications, such as oil extraction, food and cosmetic industry\cite{khan_livre}. This very common matter is also interesting as model for studying complex fluids, since their constituent can easily be observed experimentally, unlike most complex fluids. Foams exhibit a complex mechanical behavior composed of several regimes such as elasticity, plasticity or  liquid flow, depending on the foam properties and on the external constrains applied to the experimental system  \cite{weaire_book,rheo_hutzler}. Their rheological behavior is often modeled by a Herchel-Bulckley constitutive power law $\sigma=\sigma_y+k\dot{\epsilon}^{\beta}$ where $\sigma_y$ is  the yield stress, $k$ the consistency, $\dot{\epsilon}$ the strain rate and $\beta$ the power law index accounting for the foam shear thinning. The rheology of foams has been investigated theoretically \cite{rheo_raufaste,rheo_janiaud} and experimentally \cite{rheo_hohler,rheo_dollet} both in 2D and 3D configurations. Despite of the fact that the 3D cases is more realistic, local behavior is difficult to observe. Therefore in the past few years studies have been focused on 2D foams for which the bulk behavior might easily be connected to local quantities \cite{Weaire2D,Lauridsen2D}. 

The flow of 2D foams has been studied extensively in different geometries such as Couette geometries \cite{rheo_clancy,cox_couette} and linear flows \cite{force_wang,dollet_linear}. The behavior of the foam depends on the system geometry as well as on the confining boundaries. Three different basic configurations have been investigated. The bubble rafts \cite{katgert} consist of a single layer of bubbles freely floating on a surface of water. The confined bubble rafts \cite{lauridsen} are bubble rafts with a plate on top. The third configuration is the Hele-Shaw cell \cite{debregeas} with bubbles confined between two solid plates. These studies pointed out the significance of the boundaries and their influence on the bubbles flow \cite{weaire_friction,force_terriac,force_janiaud}. Bubbles in relative motion with a wall are subjected to a viscous friction proportional to Ca$^{2/3}$ where Ca is the capillary number. The power law index depends on the microscopic properties of the foam.

In the present paper, we propose to study the foam flow inside a circular Hele-Shaw cell with a unusual setup. This type of setup has already been used but for different fluids like water \cite{thoroddsen} or granular media \cite{geoffroy}. After a description of the experimental setup, we focus on the velocity profiles of the foam. Rheological tools are then introduced in order to characterize the local properties of the foam. Finally a semiempirical model is eventually proposed and validated.

\section{Experimental Setup}					

All the experiments presented below have been carried out in a circular Hele-Shaw (HS) cells made of polycarbonate plates with a radius $R=65$~mm and a thickness $e=3$~mm. The cell was half-filled with an aqueous surfactant solution composed of $94 \%$ of bi-distilled water,  $5 \%$ of glycerol and $1 \%$ of commercial dishwashing soap with low surface modulus. This solution properties are a viscosity $\eta_l\simeq 1.1~10^{-3}$~Pa.s, a surface tension $\gamma\simeq 25~10^{-3}$~N/m and a volumic mass $\rho\simeq 1.012~10^3$~kg/m$^3$. Afterwards monodisperse foam is added using a milli-fluidic T-jonction in order to fill the second half of the cell. The bubble diameter is  $D=3$~mm. The cell is then vertically fixed onto a tumbler. This configuration enables the cell to rotate around its geometrical center with an angular velocity $\omega$ ranging in $[0;1.8]$~rad/s. The direction of rotation has been arbitrarily chosen to be clockwise. The HS cell is backlighted using a circular neon light and the experiments of the foam are recorded using a high-speed video camera with a rate of $100$~fps. For some experiments a LED light is added in front of the cell to enhance the Plateau borders of the foam and to ease their detection during image processing.

\begin{figure}[h]
\begin{center}
\includegraphics[width=9cm]{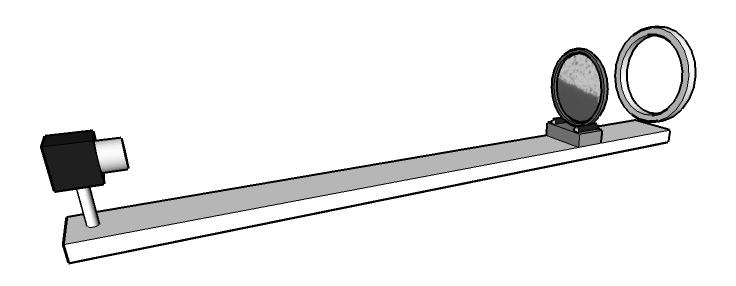}  			
\caption{Sketch of the experimental setup.}\label{fig:setup}
\end{center}
\end{figure}

When the filled cell is put on the tumbler and is left at rest the liquid inside the Plateau borders flows down under gravity. After a few minutes this flow becomes negligible, the foam has reached the hydrostatic equilibrium. The liquid weight is balanced by the capillary forces. The foam liquid fraction $\phi_l$ at gravitational equilibrium is then ruled by the equation \cite{lesmousses}:
\begin{equation}
\phi_l(y)^{-1/2}-\phi_l^{*-1/2}\simeq\frac{yD}{l_c^2},
\label{Eq:LiqFracGrav}
\end{equation}
with $y$ the vertical position with the axis origin at the foam interface, $l_c$ the capillary length and $\phi_l^*$ the liquid fraction at the interface ($y=0$) is equal to the one of a foam made of spherical bubbles.
For a cell half-filled with foam, the liquid fraction averaged over all the foam is equal to $\phi_{grav}\simeq0.01$ if $\phi_l^*=0.16$, the liquid fraction for a disordered 2D spherical bubbles foam. Later on, we will call this state the static equilibrium. In this state the foam liquid fraction decreases fast with the height. Only the first layers of bubbles are wet while the rest of the foam can easily be considered dry.

When the HS cell is set in motion, the foam previously in a static equilibrium state is rotated as a solid and the interface inclines but remains plane at first. After a relatively short time ($\sim1$~s) the bubbles start to flow upwards along the interface and along the rising edge of the cell inside the foam. The interface is then deformed into a S-shape with an amplitude increasing with the cell angular velocity. The bubbles rising from the interface add liquid to the foam and thus increase its mean liquid fraction. This rise continues until the liquid fraction is almost uniform throughout the entire foam. Nevertheless the foam on the cell rising side will remain slightly wetter than the one on the opposite side. The bubble layers at the interface are obviously also wetter. 

\begin{figure*}[htb]
\begin{center}
\includegraphics[width=18cm]{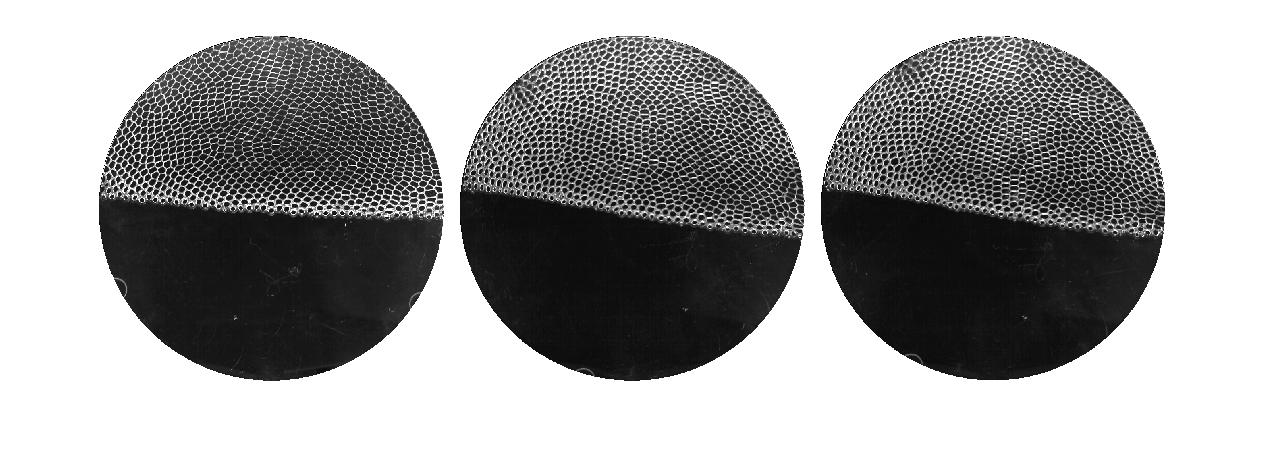} 	
\caption{Foam at the dynamic equilibrium for three different cell angular velocities $\omega$. (left) $\omega=0.085~$rad/s. (center) $\omega=0.85~$rad/s. (right) $\omega=1.7~$rad/s.}\label{fig:compV}
\end{center}
\end{figure*}

A foam and its interface stabilized in a rotating cell are in the so-called dynamic equilibrium. This state is represented in Figure~\ref{fig:compV} for three different cell angular velocities $\omega$. When $\omega$ increases the foam mean liquid fraction increases. This variation can be related to the brightness of the foam. The highest the liquid fraction the brightest the foam. A fit to the experimental data of the evolution of $\phi_l$ function of the cell velocity gives the following law:
\begin{equation}
\phi_l(\omega)=\phi_{grav}+0.03~\omega.
\label{Eq:evolphi}
\end{equation}
The liquid fraction is assumed to be uniform. Figure~\ref{fig:compV} also shows the increase of the interface slope with $\omega$. On the right-hand side picture the interface deformation is difficult to notice. But on the other pictures the cell velocity increases and the interface deformation increases. The evolution of this slope is best characterized by the angle formed by the interface center with the horizontal plane $\alpha$. This point is represented by a white dot on Figure~\ref{fig:colonnes} and corresponds to the inflection point of the interface S-shape.
 follows the experimental law
\begin{equation}
\alpha(\omega)=0.19~\omega^{0.37}
\label{Eq:evolalpha}
\end{equation}
These two equations will be useful later on.

Before each experiment the foam was left to rotate at constant velocity for a few minutes in order to reach the dynamic equilibrium, i.e. when the mean liquid fraction variations become negligible. Once the dynamic equilibrium is reached, bubble deformations and their excursions fluctuate around a mean value. 

Foams are complex fluids made of finite elastic elements. In these types of medium stationary flow results from elastic energy loads and discrete relaxations due to bubble rearrangements such as T1 events. Foams might be considered as a continuous medium whenever these bubble rearrangements do not exhibit large-scale correlations, i.e. their behavior are not dominated by T1 avalanches. These avalanches are characterized by a succession of small increases and fast decreases of locally defined quantities such as the bubble deformations. The continuous medium hypothesis has been verified by analyzing the statistical distribution of the bubble deformations fluctuations around their mean value.These distributions are well fitted by a Gaussian curve. No significant asymmetry, feature of avalanches, has been observed. This hypothesis is strong for high cell velocity with a wet foam but is a bit weaker for low velocity and dryer foams. 

\begin{figure}[h]
\begin{center}
\includegraphics[width=9cm]{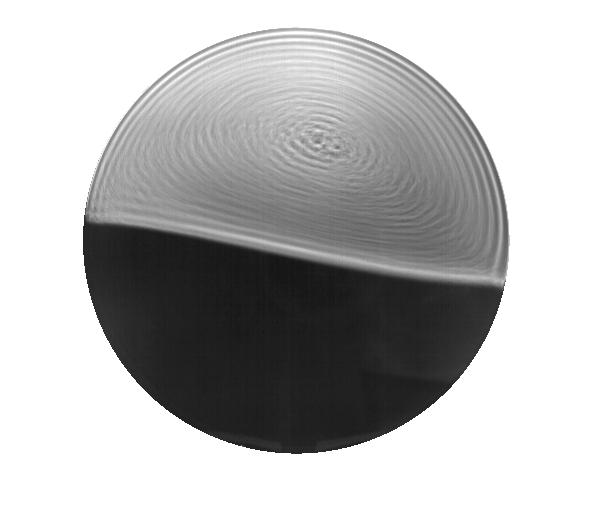}  			
\caption{Average of the pictures for a whole cell revolution with a foam at its dynamic equilibrium and $\omega=1.27$~rad/s.}\label{fig:trajectoire}
\end{center}
\end{figure}

Figure~\ref{fig:trajectoire} represents pictures of a foam averaged over a few seconds. The bubbles trajectory is really well brought out. The flow is laminar and bubbles rotates around a well defined point of null velocity $(x_0;y_0)$. This behavior is observable for all the tested cell velocities but the null velocity point position remains at the center of the foam, moving slightly with $\omega$ compared to the laboratory frame of reference. The Reynolds number Re has been computed with
\begin{equation}
\text{Re}=\frac{\rho_f V R}{\eta_{eff}}
\label{Eq:Re}
\end{equation}
where $\rho_f$ is the mass density of the foam, $V$ the linear velocity of the bubble layer in contact with the external cell wall and $\eta_{eff}$ the foam effective viscosity computed in section IV.
For all our experiments the Reynold number is small compared to unity reinforcing the idea of a laminar foam flow.

\section{Velocity Profiles}					
\begin{figure}[h]
\begin{center}
\includegraphics[width=8.95cm]{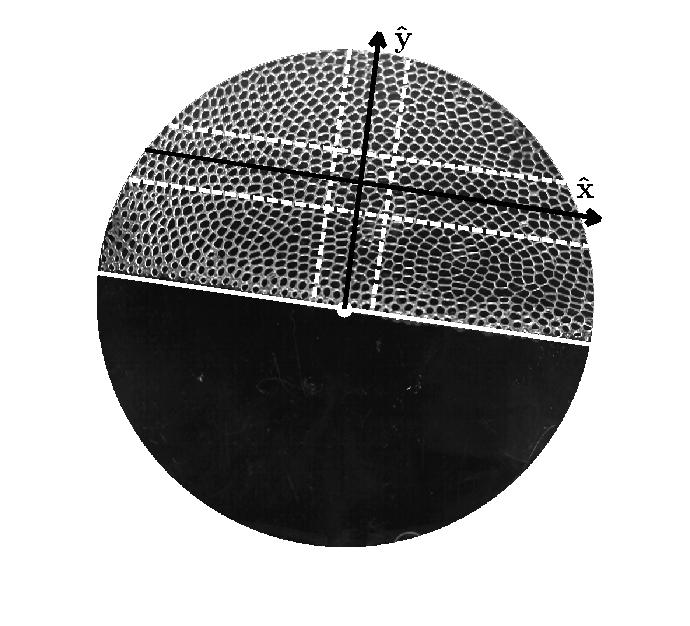}  			
\caption{Picture of a typical foam rotating at $\omega=0.87~$rad/s. The white line represents the interface approximated position. The white dotted lines represent the two representative areas limits. The white dot represents the interface center.}\label{fig:colonnes}
\end{center}
\end{figure}

To characterize the foam flow, velocity profiles have been computed along two representative lines for different cell velocities. The first line is perpendicular to the interface and represented by the y-axis on Figure~\ref{fig:colonnes} while the second one is parallel and represented on the same figure by the x-axis. Both are centered on the point $(x_0;y_0)$ around which the foam turns. 

The bubbles velocities are computed from image analysis.  The images are first treated so that the position of all bubbles is perfectly detected. Then each bubble is associated with the closest bubble on the following image. In this process we add to each bubble a vector containing their positions on all frames. Their displacement between two successive frames is then computed as well as their velocity. These informations are registered in a new data vector. To obtain the velocity profiles along the two representative line, we defined an interest area around both representative lines. These areas are approximately 5 bubble diameters wide and are centered on each representative line. They are represented on Figure~\ref{fig:colonnes} by the white dotted lines. To each bubble entering these interest areas is attached its velocity averaged over all the crossing time. This procedure is done during a long enough period of time, around $7.5\,s$,  giving us several thousands of data. Afterwards, the velocity is one more time averaged in function of the bubble position in order to keep only around fifty values of the velocity per experiment. Finally only the component perpendicular to the representative line is kept, i.e. the horizontal velocity for the vertical line and the vertical velocity for the horizontal line. Indeed along these two representative lines only the shear flow is significant. The velocity component parallel to the line is negligible.

\begin{figure}[h]
\begin{center}
\includegraphics[width=9.3cm]{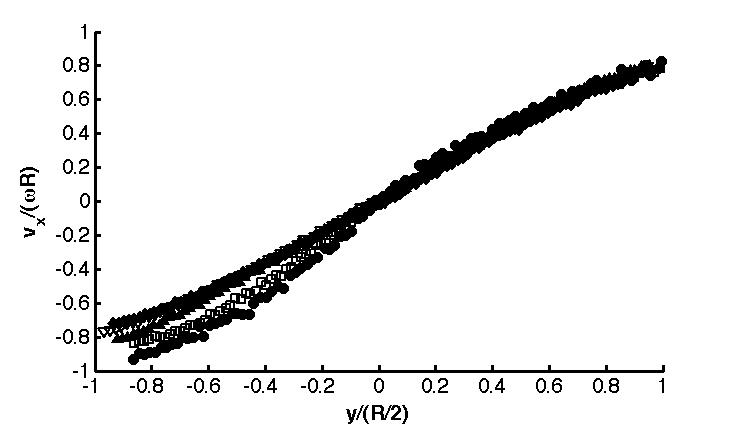}  			
\caption{Normalized velocity profiles $v_x/(\omega R)$ versus the vertical coordinate $y$ centered on the null velocity point $y_0$ and normalized by half the cell radius. Five different cell angular velocities are illustrated: $\bullet~\omega=0.087\,$rad/s $\square~\omega=0.42\,$rad/s $\blacktriangle~\omega=0.87\,$rad/s $\triangledown~\omega=1.27\,$rad/s $\blacklozenge~\omega=1.7\,$rad/s.}\label{fig:profilbbvert}
\end{center}
\end{figure}

The resulting bubble velocities along the y-axis are shown on Figure~\ref{fig:profilbbvert}. The horizontal bubble velocity $v_x$ normalized by the cell linear velocity $\omega R$ is presented as a function of the vertical bubble position $y$ normalized by half the cell radius $R$. The position is centered on the point of null velocity $(x_0;y_0)$. To ease the analysis we decided to divide the curves into two separated parts: one for the upper part of the cell, i.e. for the bubbles with a position $y>0$ and one for the lower part of the cell, i.e. $y<0$. All the data of the upper part are well normalized and collapse on a same curve with the same normalized limit velocity at the cell wall but with slightly different curvatures. The upper limit velocity is well approximated by $v_0(y=R/2)=0.8\,\omega R$ and is smaller than the cell velocity. This behavior is due to the wall slip of the bubbles. For the lower part, the data do not so well collapse. The limit velocity of the bubbles in contact with the solution $v_1$ does not evolve linearly with the cell velocity but follows the experimental law $v_1(y=-R/2)\simeq-0.6\,(\omega R)^{0.84}$. This non linear behavior might be explained by the contact with the solution interface implying a complex limit condition. The variation of the interface shape  with $\omega$ might also be playing a role. Finally, the velocity profiles curvature seems to decrease when $\omega$ increases. For both cases the profiles curvatures is positives and their variations will be analyzed later on.

\begin{figure}[h]
\begin{center}
\includegraphics[width=9cm]{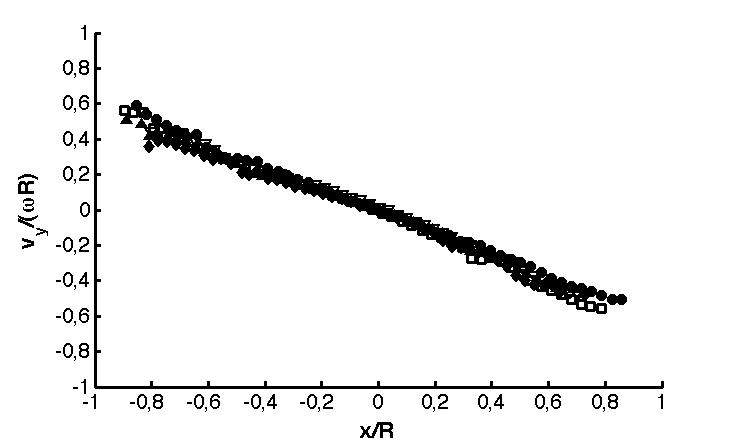}  			
\caption{Normalized parallel line velocity profiles $v_y/(\omega R)$ versus the horizontal coordinate $x$ centered on the null velocity point $x_0$ and normalized by the cell radius for five different cell angular velocities. $\bullet~\omega=0.087\,$rad/s $\square~\omega=0.42\,$rad/s $\blacktriangle~\omega=0.87\,$rad/s $\triangledown~\omega=1.27\,$rad/s $\blacklozenge~\omega=1.7\,$rad/s.}\label{fig:profilbbhoriz}
\end{center}
\end{figure}

The results for the interest area parallel to the interface are shown on Figure~\ref{fig:profilbbhoriz}. The vertical bubble velocity $v_y$ normalized by the cell linear velocity is presented as a function of the horizontal bubble position $x$ normalized by the cell radius $R$. The horizontal position is also centered on the point of null velocity. The velocity profiles are once again divided into two separated parts: one for the right part of the cell, i.e. for the bubbles with a position $x>0$ and one for the left part of the cell, i.e. $x<0$.  For both parts the profile is quite well normalized especially around the null velocity point. The profiles curvature close to the cell border seems to depend a bit on the cell angular velocity $\omega$. The limit velocity for both sides have a similar behavior and increase linearly with $\omega$.

All the profiles do not collapse perfectly. The profiles are not symmetric and do not depend only on the cell velocity. This behavior can only be understood when looking at local properties of the foam.Thus a better understanding of the external strain influence on the foam is needed.

\section{Rheology}					

To obtain the foam proposed rheological properties we followed the data analyses procedure in \cite{dollet,dollet_these}. This procedure consists in drawing a link between the local statistical behaviors and the global macroscopic properties of the foam.

The experimental pictures are first skeletonized so that all bubbles are detected and their center position recorded. On the pictures a vertex is defined as a black pixel surrounded by 3 different bubbles. They allow us to detect the bubble sides as they are delimited by two vertex with 2 common bubbles. All those data are recorded in different vectors and the pictures are not required any longer. Two neighboring bubbles have 2 common vertex and a common side. A list of bubbles with their position and the one of their closest neighbors is recorded. From this bubble list, all vectors $\boldsymbol{r}$ linking the center of two neighboring bubbles are computed. This creates a network independent of the geometry of the bubble edges which are likely distorted by the skeletonization process. Then the foam area is meshed with a circular grid made of similar triangular boxes. Each bubble is associated with a single box. The size of the boxes enables us to well capture the data field variations at a macroscopic scale with a good statistical accuracy (roughly seven bubbles per box and $750$ frames per experiment).

From the bubble center network, the texture tensor $\boldsymbol{M}$ can be computed for all the bubbles $\boldsymbol{M}=\langle\boldsymbol{
r}\otimes\boldsymbol{r}\rangle$ which is average over $750$ pictures and all vectors in the box. The texture tensor is well suited for describing the bubble deformation. It stores the deformation size, direction and amplitude but is dependent on the bubble size. A more quantitative tool for describing the bubble deformation is the statistical elastic strain tensor $\boldsymbol{U}$ defined as
\begin{equation}
\boldsymbol{U}=\frac{1}{2}\left(\text{ln}\boldsymbol{M}-\text{ln}\boldsymbol{M_0}\right).
\label{Eq:U}
\end{equation}
It compares the deformation to a reference value $\boldsymbol{M_0}$ and describes the elastic strain (relative dilatation, amplitude and direction of anisotropy). The reference texture tensor $\boldsymbol{M_0}$ described a foam under no external strain and thus without any deformation. To express this the reference state is chosen isotropic and is computed for a foam in its static equilibrium. Thus the reference foam is made of spherical undeformed bubbles but with a size that might vary from one box to another. Indeed the foam liquid fraction varies with the height in the static equilibrium. This might slightly modify the apparent bubble size.

\begin{figure}[h]
\begin{center}
\includegraphics[width=9cm]{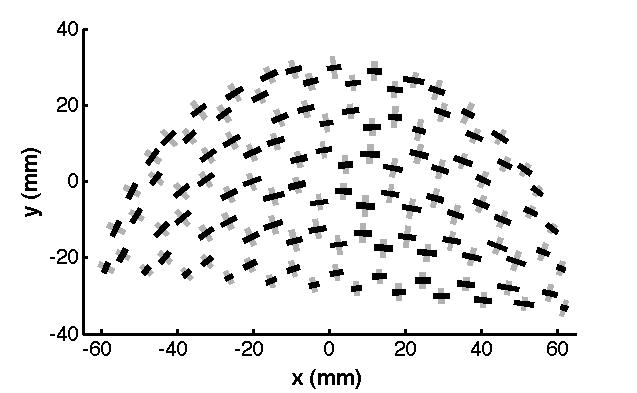}  			
\caption{Statistical elastic strain field. The grey lines correspond to the compression direction while the black ones correspond to the elongation direction. $\omega=0.25\,$rad/s clockwise.}\label{fig:mapupt}
\end{center}
\end{figure}

\begin{figure}[h]
\begin{center}
\includegraphics[width=9cm]{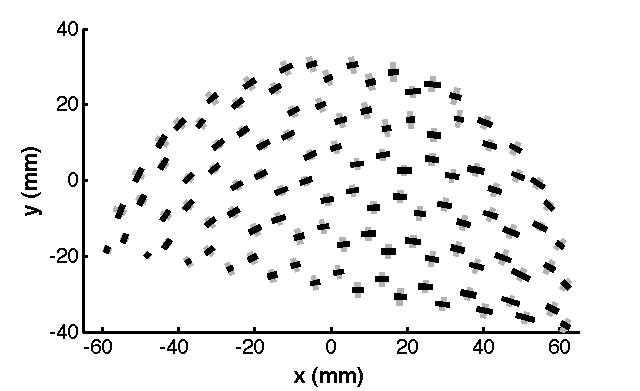}  			
\caption{Statistical elastic strain field. The grey lines correspond to the compression direction while the black ones correspond to the elongation direction. $\omega=1.7\,$rad/s clockwise.}\label{fig:mapugd}
\end{center}
\end{figure}

The tensor $\boldsymbol{U}$ extents the classical elastic strain for small deformations to larger deformations \cite{aubouy,asipauskas}. If the flow is continuous and affine the statistical elastic strain tensor can be directly equaled to the elastic strain $\boldsymbol{\epsilon}$ for a continuous media. Its trace quantifies the variation of the bubble area and thus the 2D bubble compressibility, which is small. The deviatoric elements measure the deformation due to shear. This tensor generally has a positive eigenvalue and a negative one. Figures~\ref{fig:mapupt} and~\ref{fig:mapugd} represent the statistical elastic strain tensor for the foam at two different cell angular velocities, $\omega=0.25\,$rad/s and $\omega=1.7\,$rad/s respectively. The grey lines correspond to the negative eigenvalue direction and amplitude (compression) while the black ones correspond to the positive eigenvalue (elongation).

Bubbles are stretched out in the direction of the flow. Almost all the positive eigenvalues are in the flow direction while the negative ones are perpendicular to it meaning that the compression occurs perpendicularly to the flow. The deformations are smaller along the water/foam interface. The higher fraction liquid close to the interface decreases the foam viscosity and allows the foam to flow with smaller deformations. The same behavior is observed for higher cell angular velocity values but the deformations are smaller due to the increase of the liquid fraction with $\omega$.

The velocity field is also computed by comparing two successive frames. In each virtual box meshing the foam, the bubble displacements are averaged over time. From this field the deformation rate tensor is defined as
\begin{equation}
\boldsymbol{V}=\frac{1}{2}\left(\boldsymbol{\nabla\,v}+\boldsymbol{\nabla\,v}^t\right).
\label{Eq:V}
\end{equation}
This tensor corresponds to the symmetric part of the velocity gradient tensor $\boldsymbol{\nabla v}$. It is computed in the middle of four neighbouring boxes, the two closest boxes along the vertical for the vertical gradient and the two closest boxess along horizontal axis for the horizontal one. In the continuous and affine hypothesis, the symmetrical part of the velocity gradient tensor can be directly related to the total strain rate $\boldsymbol{\dot{\epsilon}}$ \cite{graner_toolsI}. It evaluates the contribution of the external strain to the viscous stress. 

\begin{figure}[h]
\begin{center}
\includegraphics[width=9cm]{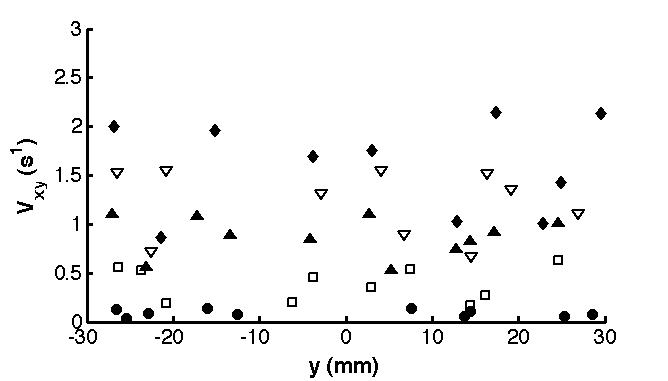}  			
\caption{The deviatoric component of the deformation rate tensor $V_{xy}$ for the vertical representative line versus the vertical position $y$. $\bullet~\omega=0.087\,$rad/s $\square~\omega=0.42\,$rad/s $\blacktriangle~\omega=0.87\,$rad/s $\triangledown~\omega=1.27\,$rad/s $\blacklozenge~\omega=1.7\,$rad/s.}\label{fig:mapvvert}
\end{center}
\end{figure}

\begin{figure}[h]
\begin{center}
\includegraphics[width=9cm]{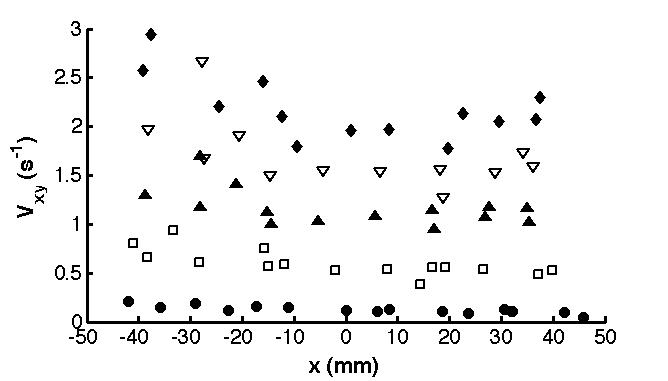}  			
\caption{The deviatoric component of the deformation rate tensor $V_{xy}$ for the horizontal representative line versus the horizontal position $x$. $\bullet~\omega=0.087\,$rad/s $\square~\omega=0.42\,$rad/s $\blacktriangle~\omega=0.87\,$rad/s $\triangledown~\omega=1.27\,$rad/s $\blacklozenge~\omega=1.7\,$rad/s.}\label{fig:mapvhoriz}
\end{center}
\end{figure}

Figures~\ref{fig:mapvvert} and~\ref{fig:mapvhoriz} show the deviatoric components of the deformation rate tensor $V_{xy}$ along the vertical and the horizontal representative lines respectively. This component will be useful for describing the shear flow of the foam along those two lines. Following the rate tensor definition, the values increase with increasing cell angular velocity $\omega$. For the vertical representative line (Figure~\ref{fig:mapvvert}) the $V_{xy}$ values at any given velocity remains approximatively constant with the vertical position $y$, revealing an almost homogenous shear along the $y$ line. For the horizontal line (Figure~\ref{fig:mapvhoriz}) the component is smaller for higher $x$ values. This behavior depicts a difference between the right-hand side and the left-hand side of the cell. On the cell left-hand side, the bubbles rising from the interface carry away some liquid such that the local liquid fraction sightly increases. This lowers the foam local viscosity and enhances the shear flow in this area. During its way to the cell right-hand side the foam drains and the liquid fraction decreases a little. This causes the foam local viscosity to increase what reduces the shear flow.

The picture analysis gives information about different phenomena occurring inside the foam. Indeed, depending on the external strain a foam can exhibit different behaviors \cite{herve, rheo_weaire}. In the elastic regime, the applied external perturbation is small enough to deform the bubbles still avoiding any topological rearrangement. This deformation is reversible. In the plastic regime the external perturbation exceeds a threshold and irreversibly deforms the foam. Bubbles undergo topological rearrangements with a very small deformation rate. At higher shear rate the foam flows like a liquid. This regime is called the viscous regime. The modeling difficulty lies in the fact that these regimes can occur separately or simultaneously. In order to model a stationary flow the stress $\boldsymbol{\sigma}$ is divided into two terms $\boldsymbol{\sigma}=\boldsymbol{\sigma_y}+\boldsymbol{\sigma_v}$. The first one is the yield stress $\boldsymbol{\sigma_y}$ corresponding to the threshold above which the foam flows. This term comes from the elastic response of foam. The tensor diagonal elements point out the foam response to compression while the deviatoric one shows the response to shear stress. Using the experimental data for the deformation the yield stress follows the law \cite{marmottant_toolsII}:
\begin{equation}
\boldsymbol{\sigma_y}=2G\boldsymbol{U_d}+K(\text{Tr}\boldsymbol{U})\boldsymbol{I},
\label{Eq:yielding}
\end{equation}

where $G$ is the elastic modulus, $K$ is the compression modulus, $\boldsymbol{I}$ is the identity tensor and $\boldsymbol{U_d}=\boldsymbol{U}-1/2(\text{Tr}\boldsymbol{U})\boldsymbol{I}$ is the deviatoric elastic strain tensor. Both the elastic and the compression moduli depend on the liquid fraction. Moreover order of magnitude estimations and experimental measurements show that $G$ is of the order of the unity while $K$ is typically $10^5$.
The second term of the total stress  $\boldsymbol{\sigma}$ is the viscous stress $\boldsymbol{\sigma_v}$. It arises from the viscous friction inside the Plateau borders and the films.This term accounts for the strain rate dependency of the foam. 

By analogy with the definition of the Newtonian liquid stress $\boldsymbol{\sigma}~=~\eta_l\boldsymbol{\dot{\epsilon}}$ \cite{lesmousses}, we define an effective viscosity for the foam as a second-order tensor
\begin{equation}
\boldsymbol{\eta_{eff}}=\frac{\boldsymbol{\sigma}}{\boldsymbol{\dot{\epsilon}}}=\frac{\boldsymbol{\sigma_y}+\boldsymbol{\sigma_v}}{\boldsymbol{\dot{\epsilon}}}.
\label{Eq:visceff}
\end{equation}

The diagonal elements are several order of magnitude higher than the deviatoric ones due to the difference between the value of the moduli $G$ and $K$. The shear effective viscosity is much lower than the compression one. The shear flow is thus highly enhanced and in the following part of the present study only the deviatoric elements of the different tensors will be considered. Moreover, the range of strain rate used in these experiments is way above the quasi-static rate and the elastic stress plays a minor role in the bubbles flow. Therefore the elastic component of the effective viscosity is neglected. This assumption will be stronger for high cell velocity values $\omega$ partly due to the decrease of the elastic modulus with increase in the liquid fraction.

\begin{figure}[h]
\begin{center}
\includegraphics[width=9cm]{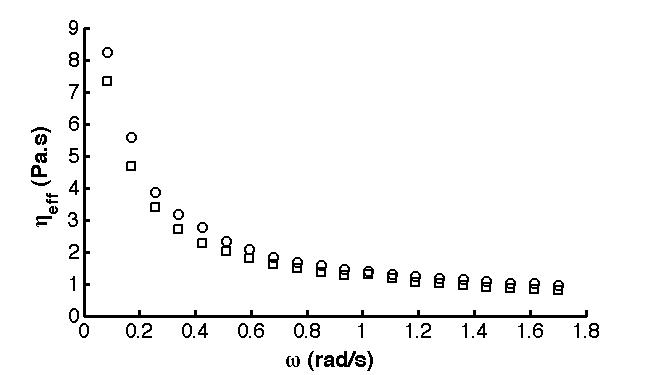}  			
\caption{Mean effective viscosities for the two representative lines in function of the cell angular velocity $\omega$. $\circ$ vertical representative line $\square$ horizontal representative line.}\label{fig:viscosite}
\end{center}
\end{figure}

The viscosity $\eta_{eff}$ is defined by the ratio of the shear viscous stress $\sigma_v$ to the shear strain rate $\dot{\epsilon}$. It is described by the law developed by Tcholakova and al. \cite{tcholakova_visc,denkov_visc}. The total shear viscous stress is considered as a superposition of the friction in foam films and the friction in the meniscus region. It follows the semiempirical formula:
\begin{equation}
\begin{split}
\eta_{eff}=\left(0.5\text{Ca}^{-0.535}\frac{(1-\phi_l)^{5/6}}{\sqrt{\phi_l}}\right)\eta_l \\
+\left(6.2\text{Ca}^{-0.3}\frac{(1-\phi_l)^{5/6}}{\phi_l^{0.2}}\right)\eta_l
\label{Eq:visceffloi}
\end{split}
\end{equation}
where $\text{Ca}=\eta_l\dot{\epsilon}D/2\gamma$ is the capillary number. The shear strain rate $\dot{\epsilon}$ is equal to the deviatoric component of the deformation rate tensor $V_{xy}$ and has been determined experimentally. The first term accounts for the friction inside the films and dominates at low Ca and/or low $\phi_l$ values, while the second term accounts for the friction inside the meniscus region and dominates at high Ca and high $\phi_l$ values. Figure~\ref{fig:viscosite} shows the mean effective viscosities for the two representative lines as a function of the cell angular velocity $\omega$. As expected both viscosities decrease with increasing cell velocity. The viscosity along the vertical line is sightly higher than the one along the horizontal line. As its value decreases with an increase in the strain rate, this behavior could have been predicted from the strain rate values on Figures~\ref{fig:mapvvert} and~\ref{fig:mapvhoriz}.

\section{Models}					

The compression deformations of the bubbles have been considered negligible due to the high value of the compression modulus $K$. Foams can then be considered as incompressible liquids.  The velocity profiles can therefore be modeled with the Navier-Stokes equations for a Newtonian incompressible liquid with a variable viscosity $\eta_{eff}$.
Incompressibility leads to:
\begin{equation}
\mathbf{\nabla}\mathbf{v}=0,
\label{Eq:incomp}
\end{equation}
While the dynamic equation is:
\begin{equation}
\frac{\partial\mathbf{v}}{\partial t}+\left(\mathbf{v}\cdot\mathbf{\nabla}\right)\mathbf{v}=-\frac{1}{\rho}\mathbf{\nabla}p+\nu\nabla^2\mathbf{v}+\mathbf{f},
\label{Eq:navierstokes}
\end{equation}
where $\mathbf{v}$ is the bubble total velocity, $\nu$ the cinematic viscosity and $\mathbf{f}$ the body forces. The variations of the effective viscosity are small at high cell angular velocity [see Figure~\ref{fig:viscosite}]. Thus the Newtonian liquid assumption will be strengthened at higher $\omega$ values.

\subsection{Vertical representative line}  

For a stationary flow along the vertical representative line without convection (small Reynolds number, cfr. Section II), Equations~\ref{Eq:incomp} and~\ref{Eq:navierstokes}  give:
\begin{equation}
\frac{\partial^2 v_x}{\partial y^2}\simeq \frac{1}{\eta_{eff}}\frac{\partial p}{\partial x},
\label{Eq:NSvertical}
\end{equation}
with $\partial p/\partial x$ approximated by $\Delta P/R$ the pressure variation over a caracteristic length $R$. This equation has been solved separately for the foam above the null velocity point $y>0$, and the lower part $y<0$. The boundary conditions used for both cases were $v=0$ at $y=0$ and the velocity of the bubble in contact with the cell border $v_0$ ($y=R/2$) and with the interface $v_1$ ($y=-R/2$), respectively. The resulting velocity profiles are
\begin{equation}
v_x^u=y\left[\frac{1}{2\eta_{eff}}\frac{\Delta P^u}{R}\left(y-\frac{R}{2}\right)+\frac{2v_0}{R}\right]\;\;\;\;\;\;\;\;\;y>0
\label{Eq:profilvert}
\end{equation}
\begin{equation}
v_x^l=y\left[\frac{1}{2\eta_{eff}}\frac{\Delta P^l}{R}\left(y+\frac{R}{2}\right)+\frac{2v_1}{R}\right]\;\;\;\;\;\;\;\;\;y<0.
\label{Eq:profilhoriz}
\end{equation}

\begin{figure}[h]
\begin{center}
\includegraphics[width=9cm]{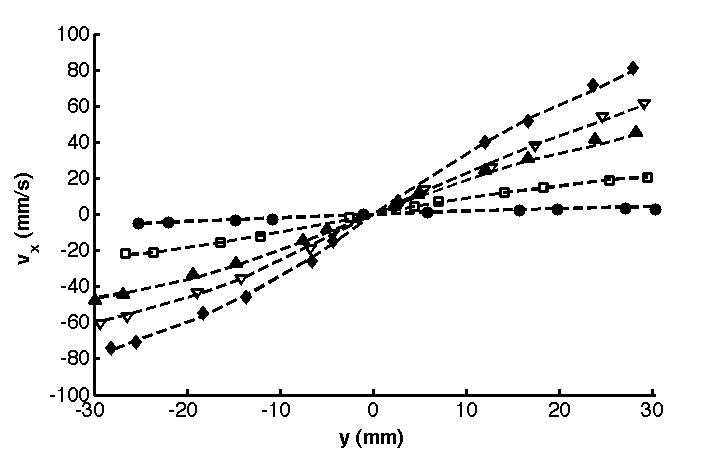} 
\caption{Vertical representative line velocity profiles $v_x$ versus the vertical coordinate $y$ centered on the nulle velocity point $y_0$ for five different cell angular velocities: $\bullet~\omega=0.087\,$rad/s $\square~\omega=0.42\,$rad/s $\blacktriangle~\omega=0.87\,$rad/s $\triangledown~\omega=1.27\,$rad/s $\blacklozenge~\omega=1.7\,$rad/s. The dashed lines represent the theoretical fitting curves.}\label{fig:profilvertfit}
\end{center}
\end{figure}

These equations have been adjusted to the experimental data with $\Delta P^u$ and $\Delta P^l$ as the free fitting parameters for both parts of the profile. Typical fits are shown on Figure~\ref{fig:profilvertfit}. The symbols stand for the velocity profiles while the dashed lines represent the model. A very nice agreement has been noticed between the model and the velocity values as well as the profile curvature of the experimental data for both sides of the profiles even for lower cell angular velocity where the hypotheses are weaker. The profile curvature points out that the pressure variations are in the flow direction for both sides of the cell leading to the requirement for the fitting parameters to be positive. 

\begin{figure}[h]
\begin{center}
\includegraphics[width=9cm]{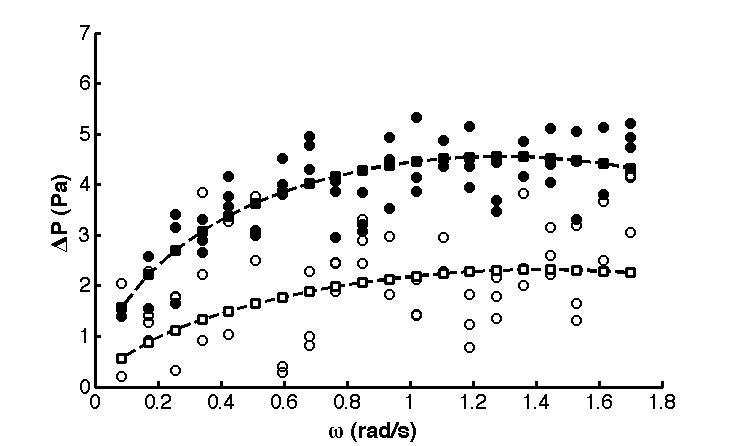} 
\caption{Horizontal pressure variations $\Delta P$ of the foam flow along the vertical representative line versus the cell angular velocity $\omega$. The symbols $\circ$ correspond to the experimental pressure variation $\Delta P^u$. The symbols $\bullet$ correspond to the experimental pressure variation $\Delta P^l$. The dashed lines represent the fitting curves for the foam upper part $\square$ and for the lower one $\blacksquare$.}\label{fig:pressionvert}
\end{center}
\end{figure}

The fitting parameters corresponding to the pressure variations are shown on Figure~\ref{fig:pressionvert} and are both positive as expected. They increase with the velocity $\omega$ and tend to saturate. The pressure variations for the foam upper part $\Delta P^u$ is significantly smaller than the lower part one. This difference was expected from the higher curvature of the lower part profiles compared to those for the upper part in Figure~\ref{fig:profilvertfit}.

While considering the Navier-Stokes equations, only the internal viscous dissipation has been taken into account so far. This dissipation is due to the relative motion of the bubbles that leads to viscous flows inside the liquid films and the Plateau borders. In a 2D foam an important ratio of the bubbles area are also in contact with the cell walls. In order to better understand the pressure variation observed in the velocity profile the external dissipation has to be considered. This dissipation is due to the the friction between bubbles and the cell walls in relative motion. Denkov et al. \cite{denkov_wall} have used the lubrication approximation to calculate the friction force between a single bubble and a wall. They have estimated the average wall stress from the relation between the foam micro-structure and its macroscopic properties. The following law was obtained for a foam with tangentially mobile interface.
\begin{equation}
\sigma_w=3C_M\left(\frac{2\gamma}{D}\right)\sqrt{1-3.2\left(\frac{1-\phi_l}{\phi_l}+7.7\right)^{-1/2}}\text{Ca}^{*2/3},
\label{Eq:wallstress}
\end{equation}
where $Ca^*$ is the capillary number defined with respect to the relative velocity of the foam and the wall $V_{wall}$ and $C_M$ is a numerical constant estimated to $3.9$ in~\cite{dollet_friction}. This law for the wall stress has been multiplied by two to account for the two walls of the Hele-Shaw cell $-\,\Delta P\simeq2\sigma_w\,-$ and has been adjusted to the pressure variations data with $C_M$ as the fitting parameter. The sliding velocity $V_{wall}$ has been experimentally measured by the difference between the wall linear velocity and the bubble velocity averaged over al the vertical positions. The bubble velocity has been approximated by $v=[v_0/(R/2)]y$ for $y>0$ and $v=[v_1/(R/2)]y$ for $y<0$. For the upper part both the walls and the foam move in the same direction leading to a relatively small sliding velocity in the flow direction. But for the lower part they move in opposite direction leading to a larger sliding velocity in the direction opposite to the flow. This easily explains the difference between the two pressure variations. The result of the fittings is represented by dashed lines on Figure~\ref{fig:pressionvert}. The pressure horizontal variations along the vertical line are well explained by the wall friction law. For both cases the fitting parameter has a value $C_M\simeq4.35$ and is similar to the values obtained in the litterature \cite{denkov_wall,dollet_friction}.

\subsection{Horizontal representative line} 

\begin{figure}[h]
\begin{center}
\includegraphics[width=9cm]{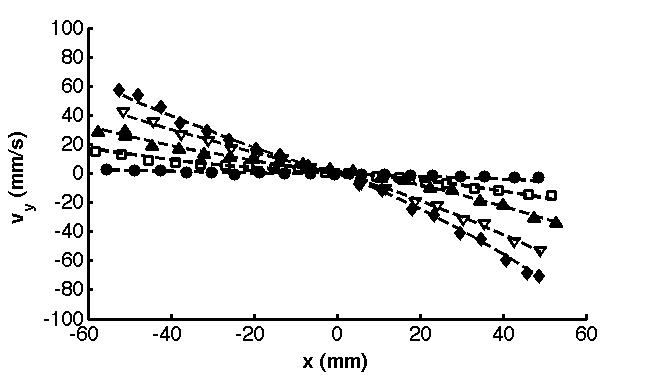}  			
\caption{Horizontal representative line velocity profiles $v_y$ versus the vertical coordinate $x$ centered on the nulle velocity point $x_0$ for five different cell angular velocities: $\bullet~\omega=0.087\,$rad/s $\square~\omega=0.42\,$rad/s $\blacktriangle~\omega=0.87\,$rad/s $\triangledown~\omega=1.27\,$rad/s $\blacklozenge~\omega=1.7\,$rad/s. The dashed lines represent the theoretical fitting curves.}\label{fig:profilhorizfit}
\end{center}
\end{figure}

The Navier-Stokes equations (Equations~\ref{Eq:incomp} and~\ref{Eq:navierstokes}) written for the velocity profiles along the horizontal representative line give the following differential equation:
\begin{equation}
\frac{\partial^2 v_y}{\partial x^2}\simeq \frac{1}{\eta_{eff}}\left(\frac{\partial p}{\partial y}+ \rho_f g\right),
\label{Eq:NShoriz}
\end{equation}
where $g$ is the acceleration of gravity and $\rho_f$ the volumic mass of the foam. The vertical pressure derivative $\partial p/\partial y$ includes two terms. The first term comes from the wall friction as for the vertical representative line. The second one takes into account the hydrostatic pressure variation due to the bubbles vertical displacement inside the foam, such that:
\begin{equation}
\frac{\partial p}{\partial y}\simeq \frac{\Delta P}{R}-\rho_f g,
\label{Eq:variationpression}
\end{equation}
where the pressure variation $\Delta P$ due to the external friction is one more time used as the free fitting parameter. Considering Equation~\ref{Eq:variationpression} the Navier-Stokes equation is solved and the resulting velocity profiles centered on the null velocity point are shown on Figure~\ref{fig:profilhorizfit}. These profiles correspond to shear flows with a pressure variation in the direction opposite to the flow for both sides of the cell. This means the pressure derivative $\partial p/\partial y$ should be negative and is ensured by the hydrostatic term in Equation~\ref{Eq:variationpression}. The curves of both sides of the cell appear to be quite similar in intensity and in curvature for a given cell angular velocity $\omega$. 

\begin{figure}[h]
\begin{center}
\includegraphics[width=9cm]{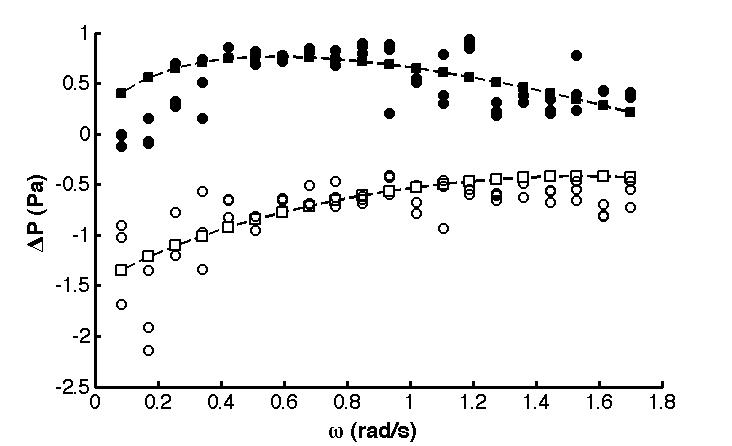} 
\caption{Vertical pressure variations $\Delta P$ of the foam flow along the horizontal representative line versus the cell angular velocity $\omega$. The symbols $\circ$ correspond to the experimental pressure variation $\Delta P^g$ on the left-hand side of the cell. The symbols $\bullet$ correspond to the experimental pressure variation $\Delta P^r$ on the right-hand side of the cell. The dashed lines represent the fitting curves for the foam right-hand part $\square$ and for the left-hand one $\blacksquare$.}\label{fig:pressionhoriz}
\end{center}
\end{figure}

The fitting parameters of the horizontal velocity profiles are represented on Figure~\ref{fig:pressionhoriz} for both sides of the cell. This parameter is the pressure variation due to the external friction and is approximated by twice the wall stress $\sigma_w$. The sliding velocity $V_{wall}$ has been estimated by the difference between the cell vertical velocity along the horizontal representative line and the linearly approximated bubbles velocity. The low $\Delta P^g$ values on the left-hand side of the cell can easily be explained in term of sliding velocity. Indeed close to the external cell wall ($y=-R/2$) the cell vertical velocity is higher than the bubble velocity leading to a sliding velocity in the direction of the flow. Closer to the null velocity point $x_0$  the situation is reversed and the sliding velocity is in the opposite direction. As $V_{wall}$ is averaged over all the left-hand part, i.e. every $x<0$ its value is small. This leads to a weak external friction. The numerical constant of the wall strain law for the pressure variation $\Delta P^g$ in Figure~\ref{fig:pressionhoriz} is close to $C_M\simeq5.4$. 

For the right-hand side of the cell the pressure variations $\Delta P^r$ are all negative. Its values are small and tend to zero when the cell velocity $\omega$ increases. The wall stress $\sigma_w$ depends on the sliding velocity of the bubbles. The bubbles flowing in the same direction than the cell walls leads to a positive sliding velocity. The wall stress is then expected to be positive and to have small values increasing with $\omega$. Its order of magnitude corresponds to the one of $\Delta P^r$ contrary to its sign and its $\omega$-dependency. Hence the wall stress is insufficient to describe the pressure variations on the right-hand side of the cell. The missing explanation may come from the local compression of the foam in the lower right corner of the cell. Indeed along the interface the cell side walls moves in the opposite direction of the foam compressing it to the right side. The descending edge ($x>0$) of the cell also pushes the foam in the right corner. The combination of these two phenomena generates an increase in the foam local pressure and influences the horizontal velocity profiles $v_y$ on the right-hand side of the cell. This need to be taken into account with an additional pressure term in the fitting equation of the pressure variations $\Delta P^r$. A negative independent term has been added to the wall stress                      leading to the equation $\Delta P^r=2\sigma_w-A$ with a numerical constant $C_M=2.89$ and $A=1.53$~Pa.

\section{Discussion}						

In this paper we used complex rheological measures to create a very simple efficient model. It led us to a good understanding of our experiment.

The foam properties and behavior depend on the cell angular velocity $\omega$. At very low velocity $\omega \rightarrow 0$, the effective viscosity of the foam tends to a very large value  $\eta_{eff}\rightarrow\infty$ (see Figure~\ref{fig:viscosite}) and the foam behaves like a solid. The external strain is below the foam yield strain $\sigma_y$ and is unable to generate any flow or any topological rearrangement. The foam is constituted only by one area behaving like a solid and the bubbles deform reversibly without any relative motion. 

\begin{figure}[h]
\begin{center}
\includegraphics[width=7.5cm]{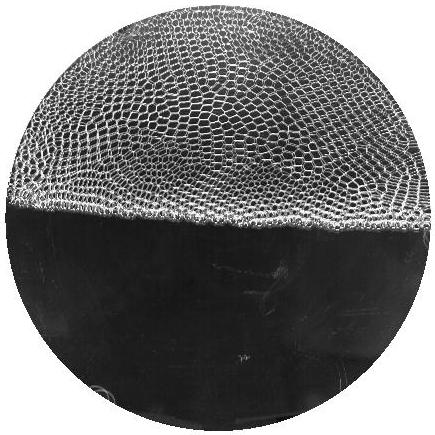} 
\caption{Average of two pictures of a typical foam rotated at $\omega=0.085$~rad/s. This figure represents the solid areas and the T1 avalanche areas.}\label{fig:avalanches}
\end{center}
\end{figure}

With slightly larger $\omega$, the effective viscosity has a finite value and the external strain is able to generate topological rearrangements. However the foam is divided into several areas where the bubbles behave like a solid and have a null relative velocity. These solid areas have a non-null relative velocity generating T1 avalanches at their boundaries. A typical foam at low $\omega$ is represented on Figure~\ref{fig:avalanches}. The largest solid area is located along the upper wall of the cell. The displacement of the bubbles are the same for almost the whole upper half of the foam along the circumference of the cell. A second smaller solid area is located along the interface. Between these two area the bubble displacements are not as well ordered. These two lines correspond to the T1 avalanche areas. Our model is based on the hypothesis of continuous medium and requires the absence of large scale correlation of T1 events. This hypothesis is not really respected at very low cell velocity.

\begin{figure}[h]
\begin{center}
\includegraphics[width=7.5cm]{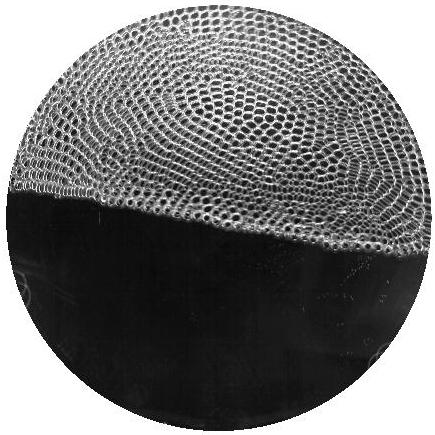} 
\caption{Average of two pictures of a typical foam rotated at $\omega=1.4$~rad/s. This figure represents the fluid-like flow of the foam.}\label{fig:continu}
\end{center}
\end{figure}

At very large cell velocity, the effective viscosity tends to a constant value $\eta_{eff}\simeq1$~Pa.s. With a viscosity independent of the shear rate, the foam behaves exactly like a Newtonian fluid. The bubble flow is fluid-like and laminar. A typical foam at high $\omega$ is represented on Figure~\ref{fig:continu}. The solid-like and T1 avalanche areas have disappeared. The bubble displacements are uniform and depend only on the bubble position compared to the foam center $(x_0,y_0)$. 

The bubble deformations at large cell velocity $\omega$ values are small. Figure~\ref{fig:mapugd} represents the statistical elastic strain tensor $\boldsymbol{U}$. The bubble deformations are smaller on the left-hand side of the cell and even smaller along the interface. On this figure, the grey line representing the compression component of $\boldsymbol{U}$ and the black line representing the elongation component of each cross have the same order of magnitude. This means the deformation of the bubbles does not modify their area. The deformations are mainly elastic and the compression is negligible. This is an experimental evidence that our model that considers only the deviatoric elements of the tensors and the shear flow of the foam is consistent. The bubble elastic deformations depend on the elastic modulus $G$ and through it, depend on the foam liquid fraction $\phi_l$. The deformations decrease with an increase in the local liquid fraction as well as the foam effective viscosity (cfr. Equation~\ref{Eq:visceff}). Therefore the foam behaves more like a fluid. The areas of largest liquid fraction are the rising side of the cell (the left-hand side) and the foam in contact with the interface. They correspond to the areas of smaller deformations on Figure~\ref{fig:mapugd}. 

Figure~\ref{fig:mapupt} represent the statistical elastic strain tensor for an intermediate cell velocity for which the model hypotheses are fully respected but useful. There is no T1 avalanche and the effective viscosity depends on the strain rate. The bubble deformations are larger than at higher velocity. They are quite uniform across the foam excepted along the interface where they are smaller. This behavior also corresponds to the local variations of the liquid fraction. Indeed for a smaller cell velocity the mean liquid fraction of the foam is smaller and its local value is always larger at the interface than for the rest of the foam.

The bubble deformations are always lined up with the wall stress (see Figures~\ref{fig:mapupt} and~\ref{fig:mapugd}). The direction of the bubble elongation is lined up with the displacement of the cell walls while the direction of the bubble compression is perpendicular. This behavior confirms the predominance of the wall stress to explain the foam flow and the pressure variations in the velocity profile. The numerical constant $C_M$ in the wall stress law (Equation~\ref{Eq:wallstress}) has different values for different velocity profiles. This behavior can easily be explained by the different approximations made on the foam properties. The most important is probably the uniform liquid fraction assumption. Indeed the liquid fraction varies by a few percents from one profile to another implying a variation in density and in effective viscosity as well.

\section{Conclusions}						

We have studied the bubbles velocity profiles of a foam inside a circular Hele-Shaw cell. This setup is different from the ordinary Couette geometries. The cell is vertical and the foam rotation center is not fixed by a cylindrical wall. Moreover the foam is in contact with a deformable liquid interface.

We have used statistical tools to determine the rheological properties of the foam. From the distances between neighbouring bubbles center we have computed the deformation tensor and the deformation rate tensor. From an Herschel-Bulckley like constitutive law the foam effective viscosity has been computed in function of the foam properties and of the external constrain. The resulting viscosity is several orders of magnitude higher than the one of the surfactant solution.

We have assumed the foam to be a Newtonian fluid but with a variable viscosity corresponding to the foam effective viscosity. The Navier-Stokes equations have been solved for our particular system and along two representative lines. The pressure variations along the foam flow have been explained in terms of wall strains. Those strains results from the viscous friction between the bubbles and the walls.

The system is a 2D foam and is entirely ruled by the viscous dissipations between bubbles and with the walls. This semi-empirical model despite its simplicity explains very well the foam profiles for all the cell angular velocity.

\acknowledgments{Part of this project has been financially supported by ... .}

\end{document}